\documentclass[preprint,5p,twocolumn,times,number]{elsarticle}

\usepackage{graphics}

\usepackage{amssymb}
\usepackage{amsthm}

\usepackage{ulem}
\usepackage{tabularx}

\begin{document}

\begin{frontmatter}



\title{New method for the time calibration of an interferometric radio antenna array}


\author[1]{F.~G.~Schr\"oder\corref{cor1}}
\ead{frank.schroeder@kit.edu}
\author[3]{T.~Asch}
\author[4]{L.~B\"ahren}
\author[1,5]{J.~Bl\"umer}
\author[1]{H.~Bozdog}
\author[4,10]{H.~Falcke} 
\author[1]{A.~Haungs}
\author[4]{A.~Horneffer}
\author[1]{T.~Huege}
\author[1]{P.~G.~Isar}
\author[3]{O.~Kr\"omer}
\author[1]{S.~Nehls}

\cortext[cor1]{Corresponding author}

\address[1]{Karlsruhe Institute of Technology (KIT), Institut f\"ur Kernphysik, 76021 Karlsruhe, Germany}
\address[3]{Karlsruhe Institute of Technology (KIT), Institut f\"ur Prozessdatenverarbeitung und Elektronik, 76021 Karlsruhe, Germany}
\address[4]{Radboud University Nijmegen, Department of Astrophysics, 6525 ED Nijmegen, The Netherlands}
\address[5]{Karlsruhe Institute of Technology (KIT), Institut f\"ur Experimentelle Kernphysik, 76021 Karlsruhe, Germany}
\address[10]{ASTRON, 7990 AA Dwingeloo, The Netherlands}

\begin{abstract}
Digital radio antenna arrays, like LOPES (LOFAR PrototypE Station), detect high-energy cosmic rays via the radio emission from atmospheric extensive air showers. 
LOPES is an array of dipole antennas placed within and triggered by the KASCADE-Grande experiment on site of the Karlsruhe Institute of Technology, Germany. 
The antennas are digitally combined to build a radio interferometer by forming a beam into the air shower arrival direction which allows measurements even at low signal-to-noise ratios in individual antennas. 
This technique requires a precise time calibration. 
A combination of several calibration steps is used to achieve the necessary timing accuracy of about 1 ns. 
The group delays of the setup are measured, the frequency dependence of these delays (dispersion) is corrected in the subsequent data analysis, and variations of the delays with time are monitored. 
We use a transmitting reference antenna, a beacon, which continuously emits sine waves at known frequencies.
Variations of the relative delays between the antennas can be detected and corrected for at each recorded event by measuring the phases at the beacon frequencies.
\end{abstract}


\end{frontmatter}

\section{Introduction}
\noindent
For the study of ultra-high energy particles from the cosmos the measurement of the radio emission from secondary particle showers generated in air or dense media is evolving as a new technique \cite{Haungs09}. First measurements of the radio emission of cosmic ray air showers had been done already in the 1960`s \cite{Allan71}, but with the analog electronics available at that time, the technique could not be competitive with traditional methods like the detection of secondary particles on ground or the measurement of fluorescence light emitted by air showers. Recently, the radio detection method experienced a revival because of the availability of fast digital electronics. Pioneering experiments like LOPES \cite{Falcke05} and CODALEMA \cite{Ardouin05} have proven that radio detection of cosmic ray air showers is possible with modern, digital antenna arrays. Due to the short duration of typically less than $100\,$ns of the air shower induced radio pulse, the experimental procedures are significantly different from those of classical radio astronomy.

The main goal of the investigations is the detailed understanding of the shower radio emission and the correlation of the measured field strengths with the primary cosmic ray characteristics. 
The sensitivity of the measurements to the direction of the shower axis, the energy and mass of the primary particle are of particular interest. 
Radio antenna arrays can derive the energy of the primary particle by measuring the amplitude of the field strength, and reconstruct the direction of the incoming primary particle by measuring pulse arrival times - with the remarkable difference to other distributed sensor networks, that with LOPES, the arrival direction is reconstructed using digital interferometry which demands a precise time calibration.
Another goal is the optimization of the hardware (antenna design and electronics) for a large scale application of the detection technique including a self-trigger mechanism for stand-alone radio operation \cite{Asch07, Berg09}.

LOPES was built as a prototype station of the astronomical radio telescope LOFAR \cite{LOFAR03, Falcke06} aiming to investigate the new detection method in detail. LOPES is a phased array of radio antennas. Featuring a precise time calibration, it can be used for interferometric measurements, e.g.~when forming a cross-correlation beam into the air shower direction \cite{Horneffer09}. Thus, LOPES is sensitive to the coherence of the radio signal emitted by air showers, allowing to perform measurements even at low signal-to-noise ratios in individual antennas.

This paper describes methods for the calibration and continuous monitoring of the timing of a radio antenna array like LOPES and shows that it is possible to achieve a timing accuracy in the order of $1\,$ns by combining these methods for such kind of arrays. 
Beside the measurement and correction of group delays and frequency dependent dispersion of the setup, we use a transmitting reference antenna, a beacon, which continuously emits sine waves at known frequencies.
This way, variations of the relative delays between the antennas can be detected and corrected for in the subsequent analysis of each recorded event by measuring the relative phases at the beacon frequencies. 
This is different from the time calibration in other experiments, like ANTARES \cite{Ageron07}, ANITA \cite{Silvestri05} and AURA \cite{Hoffman06} which determine the arrival times of pulses emitted by a beacon. In addition, AURA has the capability to measure frequency shifts of constant waves for calibration \cite{Landsman09}. The use of phase differences of a continuously emitting beacons is reported for ionospheric TEC measurements \cite{Yamamoto08}, where the measurement of phases of a beacon signal is used for atmospheric monitoring, not for time calibration. Where the individual methods described in this work are more or less standard in sensor based experiments, their combination to achieve the possibility of interferometric measurements is new and applied for the first time in LOPES.

\section{The LOPES antenna array}
\begin{figure}[t]
 \centering
 \includegraphics[width=0.99\linewidth]{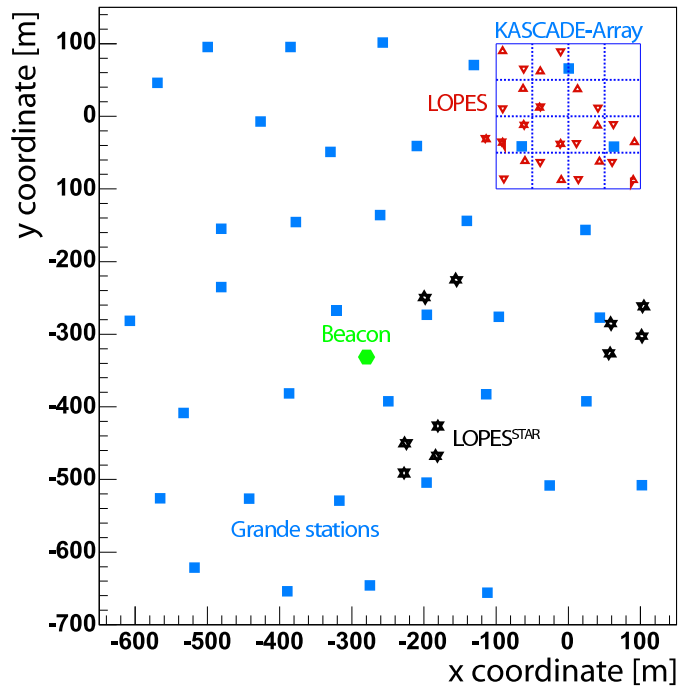}
 \caption{Current setup of the LOPES experiment which is co-located with the KASCADE-Grande experiment at Karlsruhe Institute of Technology, Germany. Upward triangles mark east-west oriented antennas, downward triangles north-south oriented antennas, respectively. A star indicates an east-west oriented and a north-south oriented antenna at the same place.}
 \label{fig_map}
\end{figure}

\noindent
The main component of LOPES consists of 30 amplitude calibrated, inverted V-shape dipole antennas \cite{Horneffer06b, Nehls08}. The antennas are placed in co-location with the particle air shower experiment KASCADE-Grande \cite{Antoni03, Navarra04} (fig.~\ref{fig_map}). 
KASCADE-Grande consists mainly of stations equipped with scintillation detectors on an area of $700\times700\,$m$^2$, where 252 stations compose the KASCADE array, and further 37 large stations the Grande array. 
Besides the 30 LOFAR-type antennas, LOPES consists also of newly designed antennas forming the LOPES\textsuperscript{STAR} array~\cite{Gemm06}. The main purpose of LOPES\textsuperscript{STAR} is to optimize the hardware for an application of this measuring technique to large scales, e.g.~at the Pierre Auger
Observatory~\cite{Abra04}. All antennas are optimized to measure in the range of $40$ to $80\,$MHz which is less polluted by strong interference than, e.g.~the FM band. 
The positions of the antennas have been determined by differential GPS measurements with a relative accuracy of a few cm.

Whenever KASCADE-Grande measures a high-energy event, a trigger signal is send to LOPES which then stores the digitally recorded radio signal as a trace of $2^{16}\,$samples with a sampling frequency of $80\,$MHz, where the trigger time is roughly in the middle of the trace. As a band-pass filter is used to restrict the frequency band to $40$ to $80\,$MHz, LOPES is operating in the second Nyquist domain and contains the complete information of the radio signal within this frequency band. Recovery of the full information is possible by an up-sampling procedure, i.e.~the correct interpolation between the sampled data points which is done by a zero-padding algorithm \cite{Kroemer08,Asch08,Nehls09}. This way, sample spacings of $0.1\,$ns can be obtained within reasonable computing time, which is considerably smaller than the uncertainties of the timing introduced by other sources (see below). Thus, the sampling rate does not contribute significantly to systematic uncertainties.

More details of the experimental set-up, the amplitude calibration, the operation, and the analysis procedures of LOPES can be found in references, e.g. \cite{Horneffer09,Nehls08}.
 
\begin{figure}[t]
 \centering
 \includegraphics[width=0.99\linewidth]{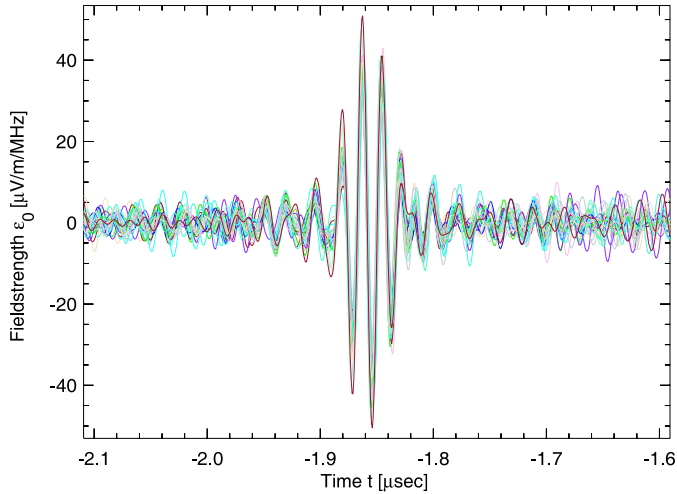}
 \caption{Calibrated, up-sampled traces of an example event, after correction for geometrical delays. The radio pulse induced by a cosmic ray air shower can clearly be distinguished from the noise, as it is (in contrast to the noise) coherently detected in all antennas.}
 \label{fig_ExampleEvent}
\end{figure}

\begin{figure}[t]
 \centering
 \includegraphics[width=0.99\linewidth]{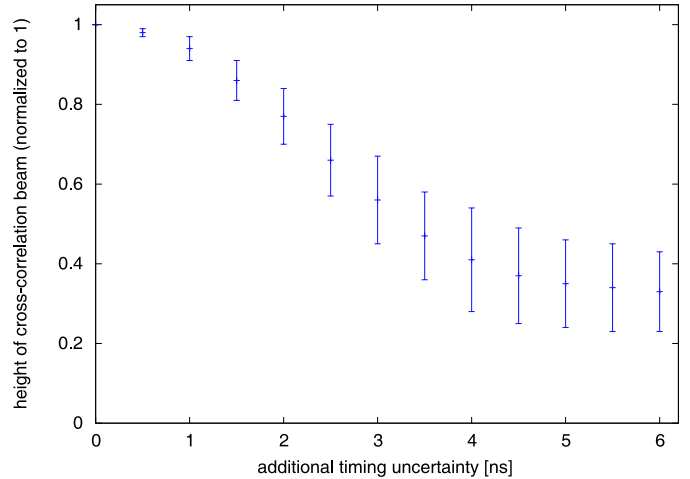}
 \caption{Influence of an additional timing uncertainty on the height of the cross-correlation beam of the example event (figure \ref{fig_ExampleEvent}): The x-axis shows the width of the Gaussian distribution of the additional timing uncertainty added to each antenna. The error bars are the RMS of 100 repetitions which have been performed for each uncertainty.}
 \label{fig_additionalUncertainty}
\end{figure}

\section{Need for a precise time calibration}
\noindent
The angular resolution, respectively source location, of LOPES is limited to about $1^{\circ}$ \cite{Nigl08a} due to the uncertainties of the emission mechanism of the radio pulse, and thus, by the uncertainties in the shape of the wave front of the radio emission. Consequently, for LOPES, improving the accuracy of the time calibration to about $1\,$ns is not expected to significantly improve the angular resolution. Instead, this good timing resolution is a necessary requirement to enable the use of LOPES as a digital radio interferometer.
Hence, this is the most important among several reasons why a precise time calibration with a relative accuracy in the order of or below $1\,$ns is desirable for a radio air shower array: 
\begin{itemize}
\item Interferometry: A timing precision which is at least an order of magnitude better than the period of the filter ringing ($\sim$\,$17\,$ns for LOPES) allows one to perform interferometric measurements if the baselines of the interferometer are adequate for the angular scale of the observed source. As the distance of the source of radio emission from cosmic ray air showers to the LOPES antenna array (several km) is much larger than the extension of the source region and the lateral extension of the array ($\sim$\,$200\,$m), the angular extension of the source is small. Hence, one expects that every antenna detects the same radio pulse just at a different time. Thus, LOPES should see coherent radio signals from air showers on the ground, which has been experimentally verified \cite{Falcke05}, and can be expemplarily seen in figure \ref{fig_ExampleEvent}. This coherence is measurable, e.g., by forming a cross-correlation beam into the air shower direction \cite{Horneffer06}, and can be used to distinguish between noise (e.g. thermal noise and noise originating from the KASCADE particle detectors) and air shower signals.

The requirement of a timing precision in the order of $1\,$ns for the interferometric cross-correlation beam analysis, can be quantitatively verified by adding an additional and random timing uncertainty to each antenna, and studying the influence on the reconstructed cross-correlation beam which is a measure for the coherence. This has been done for the example event (figure \ref{fig_ExampleEvent}) by shifting the traces of each antenna by an additional time taken from a Gaussian random distribution (see figure \ref{fig_additionalUncertainty}). The height of the cross-correlation beam decreases significantly when the added uncertainty is larger than $1\,$ns. For uncertainties $\gtrsim 5\,$ns the height is not reduced further, as the analysis always finds  a random correlation between some antennas. As most of the LOPES events are closer to the noise than the shown example, reconstructing the cross-correlation beam correctly is important, because a reduced height can lead to a signal-to-noise ratio below the detection threshold.
\item Polarization studies: Different models for the radio emission of air showers can, among others, also be tested by their predictions on the polarization of the radio signal (e.g., the geo-synchrotron model \cite{Falcke03, Huege07} predicts predominantly linear polarization of the electric field in a direction depending on the geometry of the air shower \cite{Huege05}). The capability of any antenna array to reconstruct the time dependence of the polarization vector at each antenna position, and thus, to distinguish between linearly and circularly polarized signals, depends strongly on the relative timing accuracy between the different polarization channels of each antenna.
\item Lateral distribution of arrival times: According to simulations, the lateral distribution of the pulse arrival times should contain information about the mass of the primary cosmic ray particle \cite{Lafebre09}. Only a precise relative timing, even between distant antennas ($\sim$\,$200\,$m for LOPES), can enable us to reveal this information, and to measure the shape of the radio wave front in detail.
\end{itemize}

\noindent
As stable clocks for the DAQ electronics and the trigger signal of LOPES are distributed via cables, the time calibration is basically reduced to the measurement of the electronics and cable group delays, their dependence on the frequency (dispersion), and their variations with time. Originally, the delays were measured with the radio emission from solar burst events, and their variations were monitored by measuring the phase of the carriers of a television transmitter \cite{Falcke05, Horneffer06b}.

Meanwhile, we have developed new methods for the time calibration which do not depend on external sources out of our control. Namely, we measure the delays with a reference pulse emitted at a known time, correct for the dispersion of the analog electronics and have set up an emitting antenna (beacon) which continuously transmits two narrow band reference signals to monitor variations of the delays with time.

These three methods for calibration and monitoring of the timing are combined to achieve a timing accuracy in the order of $1\,$ns for each event measured with LOPES. Nevertheless, these methods are in principle independent from each other, and for other experiments one might, e.g., determine delays by another method, but still use the beacon method to continuously monitor the relative timing.

\begin{figure}[t]
 \centering
 \includegraphics[width=0.99\linewidth]{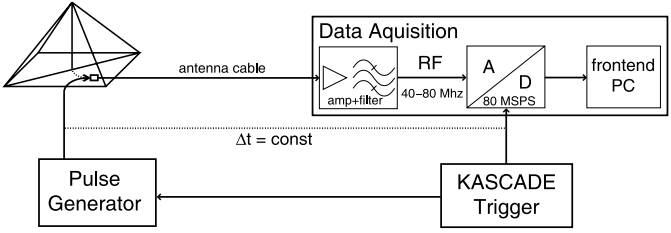}
 \caption{Setup for the delay measurements: the coaxial cable from the LOPES antenna is disconnected from the antenna and connected to a cable from the pulse generator instead which is triggered simultaneously with the DAQ electronics.}
 \label{fig_delaySetup}
\end{figure}

\section{Delay measurements} \label{sec_delay}
\noindent
For LOPES, as a digital radio interferometer, mainly the relative timing between the different antennas is of importance, and the absolute event time has to be known only roughly to combine the LOPES events with the corresponding KASCADE-Grande events. Thus, the determination of the pulse arrival times at each antenna, and therefore the measurement of the delays, is most important on a relative basis. Hereby, the delay of each channel (antenna and its analog electronics) is different, e.g., because different cable lengths are used.

We define the absolute delay $\tau$ of a channel as the time between the arrival time $t_0$ of a radio pulse at an antenna and the time  $t_{\mathrm{t}}$ when it appears in the digitally measured trace: $\tau = t_{\mathrm{t}} - t_0$. The more important relative delay $\Delta\tau_\mathrm{m, n}$ between two antennas $\mathrm{m}$ and $\mathrm{n}$ is the difference between the absolute delays of these antennas: $\Delta\tau_\mathrm{m, n} = \tau_\mathrm{m} - \tau_\mathrm{n}$.

Using solar bursts all relative delays $\Delta\tau_\mathrm{m, n}$ could be determined directly. Measuring the delays with respect to a common reference time $t_\mathrm{ref}$ is equivalent if the difference $t_\mathrm{ref} - t_0 = \mathrm{const}$ is the same for all antennas. These delays $\tilde{\tau}$ measured with respect to $t_\mathrm{ref}$ are related to the absolute delays by $\tilde{\tau} = t_\mathrm{t} - t_{\mathrm{ref}} = \tau - (t_\mathrm{ref} - t_0)$, and the relative delays can be easily derived from the measured delays $\tilde{\tau}$ by $\Delta\tau_\mathrm{m, n} = \tau_\mathrm{m} - \tau_\mathrm{n} = \tilde{\tau}_\mathrm{m} - \tilde{\tau}_\mathrm{n}$.

For each antenna the measurement of the delay $\tilde{\tau}$ is performed as follows: We disconnect the cable from the antenna and connect it to a pulse generator instead, which emits a short calibration pulse at a fixed time after a normal KASCADE-Grande trigger (fig.~\ref{fig_delaySetup}). As reference time $t_\mathrm{ref}$ we define the zero point of the LOPES trace (i.e.~$t_\mathrm{ref} = 0$), which is determined by the KASCADE-Grande trigger, because it starts the LOPES read out. As it simultaneously triggers the pulse generator of the delay measurement, the condition $t_\mathrm{ref} - t_0 = \mathrm{const}$ is fulfilled, and the delay  $\tilde{\tau}$ can by obtained as the arrival time $t_{\mathrm{t}}$ of the calibration pulse in the trace of the calibration event:  $\tilde{\tau} = t_{\mathrm{t}} - 0 = t_{\mathrm{t}}$.

\begin{figure}[t]
 \centering
 \includegraphics[width=0.99\linewidth]{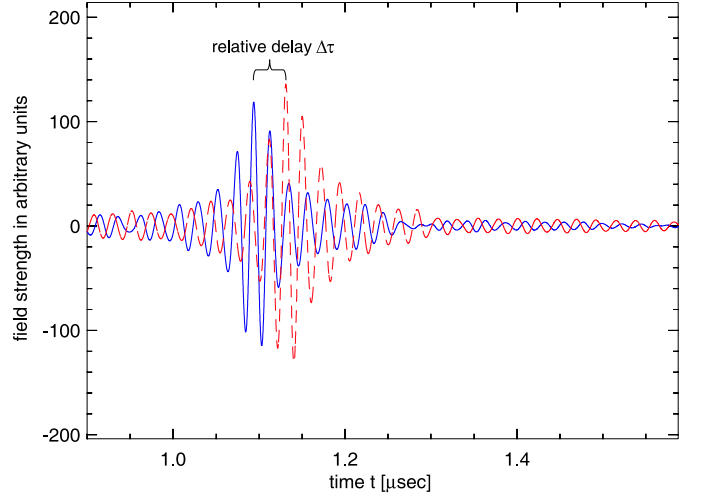}
 \caption{Example for the delay measurement: the relative delay between two antennas (solid blue and dashed red) can be measured as the difference of the times when a calibration pulse is received, which has a height of about $1\,$V and is fed directly into the antenna cables. The relative delay is mainly caused by different cable lengths.}
 \label{fig_delayExample}
\end{figure}

This pulse arrival time $t_{\mathrm{t}}$ is determined in a subsequent analysis as time of the positive maximum of the up-sampled trace (like shown in fig. \ref{fig_delayExample}).
When repeating the measurement for the same channel several times, the measured pulse arrival time $t_{\mathrm{t}}$ is stable within about one sample of the up-sampled trace (RMS of $12$ successive events $\lesssim0.1\,$ns, trace up-sampled to $(12.5/2^7 \approx 0.1)\,$ns sample spacing), if the amplitude of the calibration pulse is chosen high enough for a sufficient signal-to-noise ratio. Hence, this measurement method enables us to determine the relative delays $\Delta\tau_\mathrm{m, n}$ with a statistical error of about $\sqrt{2} \cdot 0.1\,$ns $\approx 0.15\,$ns.

Furthermore, systematic errors of the delay measurements have been studied in several ways, e.g., by repeating the measurements. Measurements of the relative delays performed on two consecutive days deviate by $(0.4\pm0.3)\,$ns from each other (mean and standard deviation of 10 measurements). As another check for systematic effects, the pulse arrival time $t_{\mathrm{t}}$ have been determined in four different ways, namely as time of the positive maximum of the trace, the negative maximum of the trace, the maximum of a Hilbert envelope of the trace, and the crossing of half height of a Hilbert envelope of the trace. The statistical error of the relative delays is about the same for each method ($\sim$\,$0.15$\,ns). But the value of the relative delays $\Delta\tau_\mathrm{m, n}$ depends on the way the pulse arrival times $t_{\mathrm{t}}$ are calculated. Only the relative delays calculated by the positive and negative maximum of the trace agree within the statistical error of about $0.15\,$ns. The relative delays calculated by the maximum of the envelope and the crossing of half height of the envelope disagree slightly with each other, and the delays calculated by the positive or negative maximum of the trace are highly inconsistent with the delays calculated by the maximum of the envelope, as they all have a statistical error of about $0.15\,$ns, but differ by up to a few nanoseconds.

\begin{figure}
 \centering
 \includegraphics[width=0.99\linewidth]{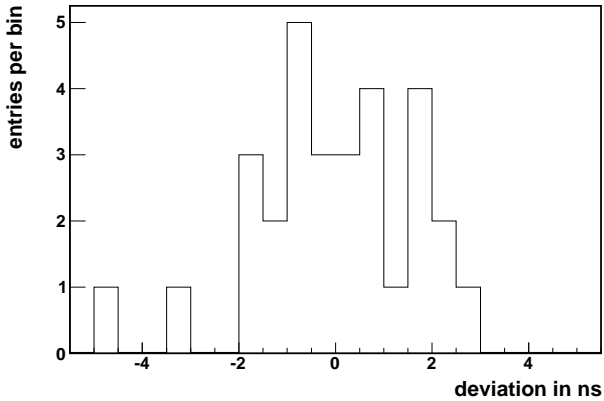}
 \caption{Deviations between delays $\tilde{\tau}$ calculated by the negative maximum of the up-sampled trace and the maximum of the envelope, mean shifted to $0$, standard deviation $= 1.7\,$ns. The histogram contains 30 deviations of one delay measurement campaign of all 30 LOPES antennas.}
 \label{fig_deviationsHistogram}
\end{figure}

Under the assumption that the electronics of all channels behaves identically, all methods for the determination of the pulse arrival times should lead to exactly the same relative delays. Hence, the explanation for the observed inconsistency is that the properties of the different channels are not exactly the same. Indeed, after correction for all measured differences, namely the amplification factor and the dispersion (see next section), the inconsistency between the delays obtained from the different methods is reduced. But still, there remains a deviation of up to a few nanoseconds for some channels, and the average deviation between the relative delays calculated by the maxima of the trace and by the envelope of the trace is of about $1.7\,$ns (see fig. \ref{fig_deviationsHistogram}). This shows the difficulty to fully correct for different channel properties. Or in other words, in designing the electronics for a new radio antenna array one has to pay attention that components are from the same batches, etc.

In the standard analysis of the shower reconstruction a cross-correlation beam is formed using the trace and not its envelope. Therefore, we have decided to use the delays calculated by the time of the positive or negative maximum of the trace of the calibration pulse. Thus, we minimize the systematic uncertainties introduced by the effect mentioned above. But still, there is another source of systematic uncertainty: The distance of the positive and negative maximum of the trace is about $9\,$ns because the response of the bandpass filter causes an oscillation with the center frequency of the used band. This oscillation and, thus, the distance between positive and negative maximum and the resulting relative delays depend only little ($\sim$\,$0.5\,$ns) on the shape of the calibration pulse. This could translate into a systematic uncertainty in the same order when determining pulse arrival times, if the pulse shape of cosmic ray radio pulses changes with lateral distance, as it is predicted by simulations \cite{Huege07}.

Another check for systematic errors was to shift the emission time of the calibration pulse by integer and non-integer multiples of the sampling clock, and no effect on the relative delays has been observed. This proves that up-sampling works reliably, and that the determination of the arrival time of radio pulses does not depend on how these pulses arrive relative to the original sampling clock. Consequently, neither up-sampling nor the original sampling rate of $12.5\,$ns do introduce any significant systematic errors.

Summarizing, the total error on the relative delays $\Delta\tau_\mathrm{m, n}$ is below $1\,$ns for the standard cross-correlation beam analysis, which is more than sufficient for interferometric measurements with LOPES. For other analysis methods, like a lateral distribution of pulse arrival times, the total error will be higher, due to the inconsistency of the different ways of calculating the pulse arrival time. In such a case the uncertainty is estimated to be in the order of $2\,$ns.

The relative delays obtained by the described method are consistent with those determined earlier by solar burst measurements. The new method, however, has two fundamental advantages compared to using astronomical sources: The resulting delays do not contain any systematic uncertainty related to the errors of the measurement of the antenna positions, and the delay calibration can be done at any time. For LOPES, this is especially important because, due to the high noise level in Karlsruhe, solar bursts are the only astronomical source visible, and thus, continuously emitting astronomical sources are not available for calibration. The described method for delay measurement is repeated roughly once per year or whenever any changes in the experimental setup require it.

\begin{figure*}[t]
  \centering
  \includegraphics[width=0.92\columnwidth]{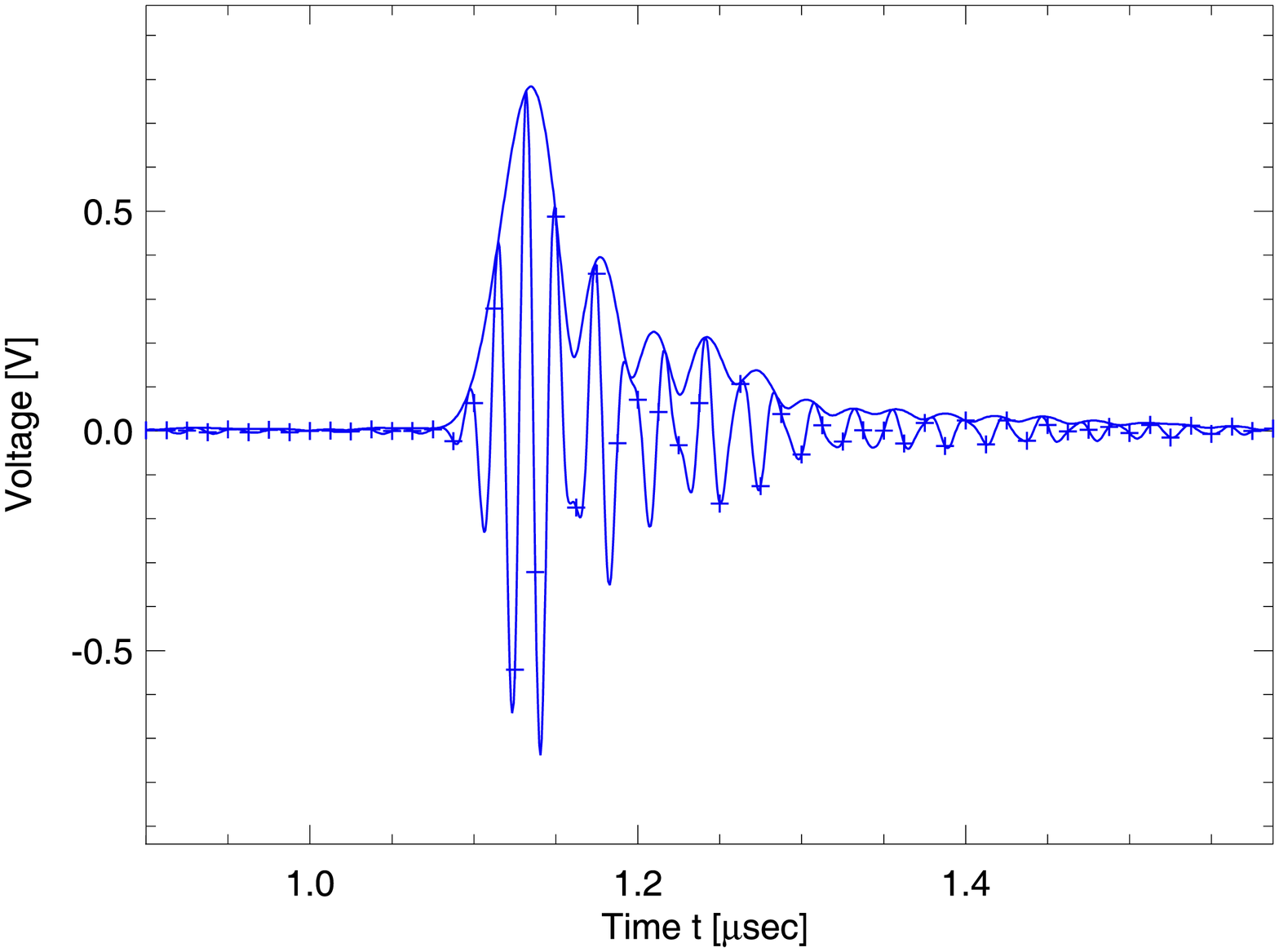}
  \hskip 0.13\columnwidth
  \includegraphics[width=0.92\columnwidth]{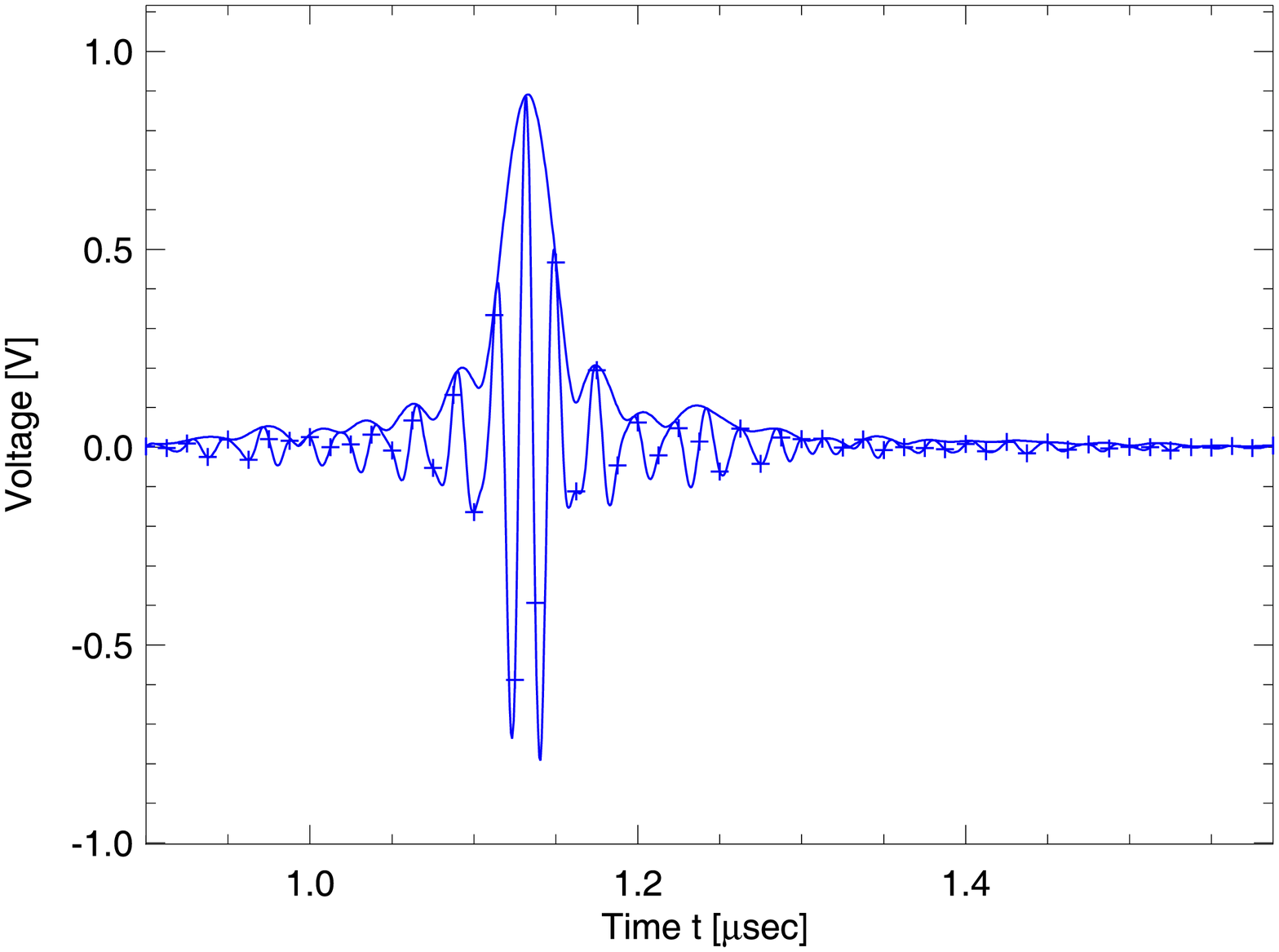}
  \caption{The pulse distortion by the dispersion of the band-pass filter is visible in the measurement of a short test pulse. Crosses indicate the sampled data points. The lines correspond to the up-sampled signal and a Hilbert envelope of the up-sampled signal. In the raw data (left), the pulse is partially delayed by more than $100\,$ns. After correcting for the dispersion in the analysis (right), the pulse becomes symmetrical and its height and width change by about $10\,$\%.}
   \label{fig_dispersion}
 \end{figure*}

\section{Dispersion} \label{dispersion_section}
\noindent
Dispersion is the frequency dependence of the group velocity, respectively of the group delay of a system. In case of dispersion, waves at different frequencies propagate with different speeds, leading to a linear distortion of broad band radio pulses. For LOPES the dispersion of the analog electronics (which is mainly caused by the band-pass filters) has been measured with a network vector analyzer. Hence, the dispersion of the filter can be removed in the subsequent analysis by multiplying the appropriate phase corrections to the frequency spectrum of any recorded data.

The effect of the dispersion has been studied with test pulses from a pulse generator which has been connected to the analog electronics instead of the antenna, like for the delay measurements. Different shapes of test pulses have been examined, and one example is shown in figure~\ref{fig_dispersion} before (left) and after correction for the dispersion (right). For most pulse shapes the dispersion leads to a change in amplitude and FWHM of an Hilbert envelope of the up-sampled field strength trace of about $10\,$\%. As the influence of the filter dispersion is largest close to the edges of the frequency band, the mentioned distortion effects can be reduced from about ten to a few percent, when using the sub-band from $43$ to $74\,$MHz, only. For radio experiments with unknown dispersion such a selection of an inner sub-band would be a possibility to reduce systematic uncertainties originating from pulse distortions.

Because the radio pulses from real cosmic ray events are similar to the used test pulses (at least within the used frequency band), distortion effects in the same order of magnitude are expected for real events (i.e.~changes of a few percent of amplitude and FWHM). In addition the pulse arrival time changes by up to a few nanoseconds, depending on how it is calculated (e.g., value at pulse maximum or at the crossing of half height). These are changes of the absolute value which have a similar effect for all channels, as equal electronics is used, and thus, the dispersion of each channel is approximately the same, which has been verified by measurement. Under the assumption that the cosmic ray radio pulse shape does not change much on the lateral extension of LOPES ($\sim$\,$200\,$m), it should be distorted by every antenna and its corresponding electronics in the same way. This means that the impact of the dispersion on the relative timing is expected to be much smaller than the observed absolute shifts of a few nanoseconds. Consequently, the dispersion of LOPES, even if not totally corrected for, should not spoil the capability to achieve a relative timing accuracy of about $1\,$ns.

As all LOPES antennas are from the same type, their dispersion is expected to affect the relative timing between the individual antennas only marginally. By this, it is acceptable that the dispersion of the LOPES antenna type is not known. It is difficult to measure, because the LOPES antenna can be used as receiver, only, and thus the two antenna method which is normally used for the determination of the dispersion, cannot be applied. In figure \ref{fig_ExampleEvent}, the traces of a real cosmic ray event are corrected for the dispersion of the filters, and the remaining pulse distortion seems to be smaller than those shown in figure \ref{fig_dispersion}, where the calibration pulse is affected by the dispersion of the filters, only. Thus, the sum of the dispersion of all other components, including the antenna, is assumed to be lower than the dispersion of the filters. Nevertheless, due to the high noise level for real events, and because the exact shape of the cosmic ray radio pulses is unknown, this can not be expressed quantitatively.

For LOPES\textsuperscript{STAR} which uses different antennas and electronics, the dispersion of the complete system has been measured \cite{Kroemer08}. It was found, that the dispersion of the cables can be neglected, but the dispersion of the antenna itself cannot. It can be of the same order of magnitude as the filter dispersion. 
For this reason, future radio experiments should aim either for antennas with low dispersion or for antennas with well-known and, thus, correctable dispersion. Correcting pulse distortions induced by the antenna dispersion is especially important for larger-scale antenna arrays, if it turns out that the cosmic ray radio pulse shape changes with lateral distance. This could also have implications for the application of interferometric analysis methods, e.g., forming a cross-correlation beam. Hence, larger antenna arrays, like AERA at the Pierre Auger Observatory \cite{Berg09} or LOFAR \cite{Falcke06}, have the opportunity to test this.

\begin{figure}
 \centering
 \includegraphics[width=0.9\linewidth,angle=180]{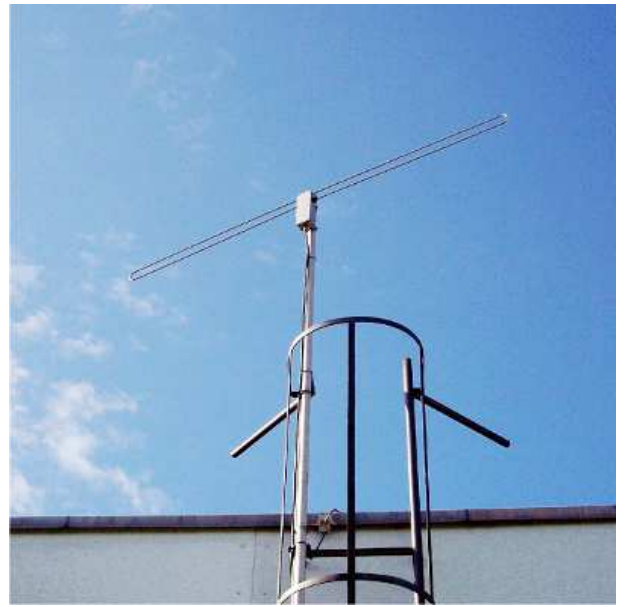}
 \caption{The dipole antenna used as beacon to monitor the timing of LOPES.}
  \label{foto}
\end{figure}

\section{Monitoring the timing with a beacon} \label{beacon_section}
\noindent
Experience with LOPES has shown that the timing is not absolutely stable. Instead, once in a while, jumps by one or two clock cycles ($12.5\,$ns) occur. In addition, small drifts or changes of the relative delays, e.g., with changing environmental temperature, cannot be excluded, as the electronics has not been designed for sub-nanosecond stability. Independent of the reasons, any changes of the timing have to be accounted for, to achieve an overall timing accuracy in the order of $1\,$ns. As the exact variations of the delay are not predictable, a continuous monitoring of the timing is needed which provides the ability to correct the timing in the subsequent analysis on an event-by-event basis.

For this monitoring we have deployed an emitting dipole antenna, a beacon, on top of a building of the Karlsruhe Institute of Technology, at about $400\,$m distance to the center of LOPES (figs.~\ref{foto} and \ref{fig_map}). This beacon permanently transmits two sine waves at constant frequencies of $63.5\,$MHz and $68.1\,$MHz (width $< 100\,$Hz) at a low power of $-21\,$dBm ($\approx 0.008\,$mW). Thus, every LOPES event contains a measurement of the phases at these frequencies, which can be obtained by a Fourier transform into the frequency domain. Any variation in the relative timing between two antennas can be detected as a variation of phase differences at each beacon frequency.

The phase of the continuous beacon signal at an antenna depends on the distance and the orientation angle of the antenna towards the beacon as well as on the delay of the corresponding channel. Let us assume for a moment, that there are two antennas at an equal distance and angle to the beacon and with an equal delay. If we consider just one beacon frequency, e.g.~$63.5\,$MHz, the two antennas would measure the same phase at this frequency (except for small deviations due to noise). Thus, a variation in the relative delay between the two antennas would immediately lead to a change of the measured phases. If, e.g.,~the relative delay shifts by $1\,$ns, the difference between the measured phases at the two antennas would be $\Delta \phi = 1\,$ns $ \cdot \, 63.5\,$MHz $ \cdot \, 360^{\circ} = 22.9^{\circ}$. Correspondingly, a measured phase difference can be converted in a shift of the relative delay.

Now let us consider the more realistic case, that we have two antennas with different angle and distance towards the beacon and different electronics and cable delays. As the distance and the effect of the antenna orientation is not precisely known (because there is no need), we expect to measure a different phase at both antennas at the beacon frequency. As long as neither the distance, nor the orientation, nor the relative delay do change, the difference between the phases measured at both antennas $\Delta \phi_\mathrm{ref}$ would be arbitrary, but constant. Thus again, changes of the relative delay can be detected as changes in the phase difference $\Delta \phi$. The only difference to the case above is, that these changes of the phase differences will happen not with respect to $0^{\circ}$, but with respect to $\Delta \phi_\mathrm{ref}$.

The important point is to define $\Delta \phi_\mathrm{ref}$ for each antenna with respect to a fixed antenna as reference (arbitrary choice) and each beacon frequency when the delay is exactly known. Therefore, we determine $\Delta \phi_\mathrm{ref}$ as an average of the events taken at the time when we do the delay calibration described in section \ref{sec_delay}. This way we can monitor and subsequently correct any variation in the timing back to the values obtained in the delay calibration.

The limitation of the accuracy of the measurement of the phase differences is given by the noise and by systematic effects. The noise of the phase measurement depends (within reasonable limits) on the signal-to-noise ratio of the beacon signal, where the amplitude of the beacon emission can be chosen such that a sufficient accuracy is achieved. In case of LOPES we have chosen to emit each frequency at $-21\,$dBm. The noise of the phase measurement has been determined by the jitter of the phase differences in successive events and corresponds to an accuracy in the order of $\sim$\,$0.3\,$ns. Aside from that, the additional noise introduced by the beacon signal to the data is negligible, as the cosmic ray radio pulses are broad band and extend over the entire frequency spectrum. On the other hand the beacon signal is visible only in a few fixed and defined frequency bins and can be suppressed by artificially reducing the amplitude at these bins in the data analysis or in the hardware of the trigger logic if a radio self-trigger system is applied.

\begin{figure}[t]
  \centering
  \includegraphics[width=0.99\linewidth]{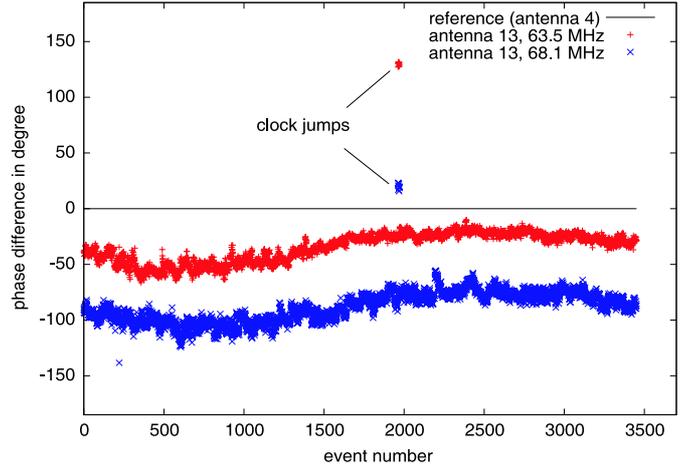}
  \caption{Phase differences between two antennas at both beacon frequencies for the first ten events per day for a whole year (May 2008 - May 2009), excluding a few days of down time. The absolute value of the phase differences is not meaningful. Though, the changes are, which amount at both frequencies to about $1.5\,$ns ($\sim$\,$35^{\circ}$) between summer and winter. Details see text.}
  \label{fig_phasediffs}
\end{figure}

\begin{table*}
\centering
\begin{minipage}{0.99\linewidth}
\renewcommand{\thefootnote}{\textit{\alph{footnote}}}
\caption{Summary of uncertainties of the relative timing and their relevance for the interferometric cross-correlation beam (CC-beam) analysis} \label{tab_summary}
\begin{tabular}{lcccc} 
Effect                                               & amount [ns] & reducible by & to [ns]\footnotemark[1] & relevant for CC-beam\\
\hline
ADC sampling frequency                               & $12.5$      & up-sampling&$\lesssim0.1$&no\\
\hline
measurement of relative delays:                      &&&&\\
~~\textbullet~ repetition on consecutive days        & \phantom{\footnotemark[2]}$0.4$\footnotemark[2]&& $0.4$ & yes\\
~~\textbullet~ different shapes of calibration pulse & $0.5$       && $0.5$ & \phantom{\footnotemark[3]}partially\footnotemark[3]\\
~~\textbullet~ method of pulse time determination    & $1.7$       && $1.7$ & no\\
\hline
remaining dispersion after correction                &&&&\\
for known properties of the filter                   & \phantom{\footnotemark[4]}$\lesssim 1$\footnotemark[4] && $\lesssim 1$ & \phantom{\footnotemark[3]}partially\footnotemark[3]\\
\hline
variations of the timing with time:                  &&&&\\
~~\textbullet~ occasional clock jumps                & $1$ or $2$ samples & beacon & 0 & (yes)\\
                                                     & ($12.5$ or $25$\, ns)&&&\\
~~\textbullet~ drifts (e.g., summer vs. winter)      & up to $1.5$ & beacon & \phantom{\footnotemark[5]}$\sim0.7$\footnotemark[5]& yes\\
\hline
total uncertainty (quadratic sum)                    & up to $\sim28$&& $\sim2.0$ & no\\ 
total uncertainty for cross-correlation beam         &&& ~\phantom{\footnotemark[3]}$0.8$--$1.4$\footnotemark[3] & yes\\ 

\end{tabular}
\footnotetext[1]{The uncertainties due to the delay measurements and due to dispersion are not further reduced.}
\footnotetext[2]{contains statistical error of $0.15\,$ns.}
\footnotetext[3]{depends on how much the shape of the air shower radio pulse changes with lateral distance.}
\footnotetext[4]{The exact amount is unknown, but assumed to be significantly smaller than a few nanoseconds (see section \ref{dispersion_section}).}
\footnotetext[5]{value determined from cross-check of beacon correction with delay measurements (see section \ref{beacon_section}):}
\footnotetext{The main contribution is a deviation of $0.6$\,ns between the results of both beacon frequencies which contains a statistical error of $0.3\,$ns for each frequency.}
\end{minipage}
\end{table*}

In figure \ref{fig_phasediffs}, the phase differences at both beacon frequencies between two LOPES antennas are shown for the first ten events of each day for one year. An annual drift of the phase differences which corresponds to about $1.5\,$ns ($\sim$\,$35^{\circ}$) can be seen consistently at both frequencies. The reason for this annual drift is not definitely known, but might be due to environmental effects, in particular changing temperature, as the effect is largest in summer and winter. Also a jump in the timing of two clock cycles ($25\,$ns)  is visible which occurs during one day. Here it becomes obvious, that at least two beacon frequencies are needed, as changes in the timing larger than half a period ($\sim$\,$9\,$ns) could otherwise not be detected unambiguously. A consistency check between the results at both frequencies is also necessary to identify a few noisy events (like the outlier in the bottom left corner of figure \ref{fig_phasediffs}), for which the beacon correction of the timing cannot be performed.

When inspecting the data carefully, some features in the plot of the phase differences can be seen which do not occur simultaneously at both frequencies - contrary to the visible general drift. These features are due to systematic effects and in principle decrease the achievable timing accuracy. Possible reasons for systematic effects are changes in the emitted beacon signal (e.g., if the frequency generation is not absolutely stable), changes in the propagation of the signal from the beacon to the LOPES antennas (e.g., due to different atmospheric or ground properties), and non-random (e.g., human-made) noise at the beacon frequencies. As the scale of the observed features is significantly smaller than $1\,$ns, they have not been investigated in detail, and should not limit the ability of the beacon method to achieve a timing accuracy of $1\,$ns.

As a cross-check, the changes of the delays between two dates roughly one year apart, have been measured with the method described in section \ref{sec_delay}, and compared to the changes of the beacon phase differences between the same two dates. The relative delays measured with the method of section \ref{sec_delay} changed by $(0.6 \pm 0.4)\,$ns between the two dates (mean and standard deviation of the absolute change of all 30 antennas). This itself is not unexpected as the electronics was not designed to be stable on a sub-nanosecond level. Comparing these changes of the delays measured with the method of section \ref{sec_delay}, with the changes observed by the beacon, reveals some systematic effects. In the ideal case, the phase differences at both beacon frequencies should change by exactly the amount corresponding to the changes of the delays. In reality, the changes observed at the two beacon frequencies are not totally equal, but the changes of the phase differences at the first beacon frequency and the second beacon frequency differ by $(0.6 \pm 0.3)\,$ns. This is larger than the statistical error which is about $0.3\,$ns (see above). The changes observed at both beacon frequencies have been averaged, to check, if they are consistent with the changes of the delays measured with the method of section \ref{sec_delay}: the changes determined by both methods deviate by $(0.7 \pm 0.5)\,$ns from each other (average of the individual deviations of all antennas).

Hence, systematic effects on the beacon signal seem to play a role, and it cannot be excluded that the observed drifts of the phase differences are - at least partly - not due to drifts of the electronics or cable delays, but due to these systematic effects. Nevertheless, this does not undermine the ability of the beacon method to monitor and correct real changes in the timing, like the described clock jumps, and to provide for each event a timing accuracy in the order of $1\,$ns which is required for digital radio interferometry.

Finally, a beacon cannot only be used to monitor the timing of an antenna array, but is valuable to check the health of the experimental setup in general. As it provides a defined reference signal visible in each event, most possible failures of the antennas or the electronics are detectable by monitoring the beacon signal. For example, we have been able to exactly find the date when we accidentally switched the cables of the two polarization channels of one antenna, by investigating the phase differences at the beacon frequencies between these channels.

\section{Conclusion}
\noindent
The methods described for the time calibration of LOPES are especially useful for radio antenna arrays in a noisy environment, where the calibration with astronomical sources is not possible. They allow the determination of the electronics and cable delays with a very high precision, which can in principle be below $0.5$\,ns. Systematic effects, however, limit the actual achieved accuracy of the delay measurement to below $1$\,ns for our standard, interferometric cross-correlation beam analysis and to about $2\,$ns for the direct measurement of pulse arrival times. In addition, the dispersion of the electronics has been measured and is taken into account in the analysis of cosmic ray air shower radio pulses, to avoid systematic uncertainties in the pulse height which can be up to $10\,$\%.

Furthermore, we continuously monitor any variation of the timing with narrow band reference signals from a beacon, thus achieving an overall timing accuracy in the order of $1\,$ns for the cross-correlation beam analysis (see table \ref{tab_summary}). This way the nanosecond time resolution required for digital radio interferometry is achieved for each event, and the phased antenna array LOPES can be used as a digital interferometer which is sensitive to the coherence of the air shower radio emission.

Finally, monitoring of the timing with a beacon is an interesting feature for any radio antenna array. As in principle the phase differences at the beacon frequencies are sensitive to any variation of the relative timing, even the timing accuracy of antenna arrays without stable clocks should be improvable to about $1\,$ns. Hence, a beacon should provide any radio experiment in the MHz regime with the capability to do interferometric measurements. For example, the application of a beacon and the possibility of interferometric measurements of the cosmic ray air shower radio pulses with larger arrays is presently investigated at the newly developed antenna array AERA at the Pierre Auger Observatory. Also LOFAR will apply the described methods for time calibration, and observe the radio emission of cosmic ray air showers with a much denser array.

\section*{Acknowledgments}
\noindent
The authors would like to thank the technical staff of the Karlsruhe institutes for their help. 
Sincere thanks to the entire LOPES and KASCADE-Grande collaborations for providing the working environment for these studies.

\end{document}